\documentclass[preprint,showpacs,showkeys,preprintnumbers,amsmath,amssymb]{revtex4}

\usepackage{setspace}
\usepackage{graphicx}
\usepackage{dcolumn}
\usepackage{bm}

\begin{document}

\title{A Particle-Particle, Particle-Density (P$^3$D) algorithm for the calculation of
electrostatic interactions of particles with slab-like geometry}

\author{S. Alireza Ghasemi}
 \email{Alireza.Ghasemi@unibas.ch}
\author{Alexey Neelov}
\author{Stefan Goedecker}
 \homepage{http://pages.unibas.ch/comphys/comphys/}
\affiliation{
Condensed Matter Theory Group,
Department of Physics \& Astronomy,
Klingelbergstrasse 82,
Basel 4056,
Switzerland
}
\date{\today}

\begin{abstract}
We present a fast and accurate method to calculate the electrostatic energy and 
forces of interacting particles with 
the boundary conditions appropriate to surfaces, i.e 
periodic in the two directions parallel to the surface and 
free in the perpendicular direction. 
In the spirit of the Ewald method the problem is divided into a short range and long range 
part. The charge density responsible for the long range part is represented  
by plane waves in the periodic directions 
and by finite elements in the non-periodic direction. 
Our method has computational complexity of 
$\mathcal{O}(N_g \log(N_g))$ with a very small prefactor, where 
$N_g$ is the number of grid points. 
\end{abstract}

\pacs{Valid PACS appear here}

\keywords{Finite Element, Slab-Like System, Poisson Solver.}
\maketitle

\section{Introduction:}

Simulations of systems with slab-like geometries are of great importance.
Problems involving 
surfaces, interfaces, tip-surface interaction in scanning probe microscopy simulations, 
electrolytes trapped between two plates, thin films of ferrofluids, etc. all fall into this category.
Calculating the Coulomb interactions in such setting is a major challenge.
With free boundary condition (i.e. the potential tends to zero at infinity)
the scaling of the trivial direct summation is $\mathcal{O}(N^2)$ where $N$ is the 
number of particles. In the case of 2D periodic and 1D free (2DP1DF) boundary 
conditions (BC) the situation is even worse. In principle one would then have to include into the 
summation the interations with all the periodic images in the two periodic directions. 

Algorithms such as Ewald-based methods \cite{cite11}, fast multipole 
methods(FMM)\cite{cite10}, P$^3$M method\cite{cite03}, and 
convergence factor approaches\cite{cite12,cite13,cite14} have therefore been 
generalized to 2DP1DF problems. Handling different  types of BC in 
FMM\cite{cite17} is straightforward. In addition the FMM methods have the 
ideal linear scaling. Unfortunately the prefactors in FFM methods are typically 
large and so the FMM methods are in many cases only faster than other 
methods for $N>10^6$, where $N$ is the number of particles. Another drawback of FMM that 
is important in molecular dynamics is that the approximate FMM forces are not 
analytical derivatives of the approximate energy. Therefore the energy is not 
conserved during the molecular dynamics simulation. High accuracy energy 
conservation is therefore impossible. \par

Ewald methods 
for 2DP1DF boundary conditions, called EW2D,  have been developed 
Refs. [\onlinecite{cite19,cite20,cite21}]. 
A comparison of three versions of EW2D method can be found in 
Ref.[\onlinecite{cite09}].Unfortunately, the practical 
use of the EW2D sum is hampered by the occurrence of a reciprocal 
space term. The resulting Fourier space sum does not allow 
for a product decomposition as it is done in the three-dimensional periodic Ewald method and therefore 
the method has a scaling of $\mathcal{O}(N^2)$. 
In 2002 Arnold and Holm developed 
MMM2D\cite{cite01}(MMM with 2DP1DF BC), which is found to be the best in terms 
of accuracy.  Another advantage of this method is that it has ``{\it a priori}'' error estimates.
However, because of its $\mathcal{O}(N^\frac{5}{3})$ scaling it is only suitable for a 
small number of atoms. 

A rather simple approach is to use the standard three dimensional periodic Ewald method (EW3D) also 
for 2DP1DF boundary conditions. Spohr showed that the 
regular EW3D method almost 
reproduces the EW2D results\cite{cite02}, provided that the box length in 
the non-periodic direction is about five times larger than those 
in the periodic directions and that there is empty space of 
sufficient thickness in the basic periodic box to dampen out 
the inter-slab interactions. There are also methods with correction terms to 
make the 3D periodical scheme applicable to the 2DP1DF systems and resolve the problem of
slow convergence with respect to thickness, so that a medium size gap(empty space) is enough.
The EW3DC\cite{cite07,cite06} method consists of a modification of EW3D to account for the slab 
geometry and addition  of a correction term to remove the 
forces due to the net dipole of the periodically repeating slabs.
Methods with layer correction terms to eliminate the inter-slab interaction, 
in addition to the correction term responsible for net dipole, have been mixed 
with mesh-based methods, thus almost linear scaling is achieved e.g. 
EW3DLC\cite{cite15,cite16}, P$^3$MLC\cite{cite15,cite16}. The main drawback of 
these methods is that the errors in the forces on the particles near to the 
surfaces are more than in the middle. 

In this paper we present a method which fills the gap of absence of an 
efficient method for medium size systems having $10^2-10^6$ particles.
Because our method is not based on a modification of a fully periodic method, no replication 
is needed in the non-periodic direction, leading to smaller memory and CPU requirements. 
In contrast to some others, our method does not impose any restriction on the distribution 
of particles in the non-periodic direction.

\section{Coulomb Interaction for Systems with 2DP1DF BC}

Consider a system of N particles with charges $q_i$ at positions ${\bf r}_i$
in  an overall neutral and rectangular simulation box of dimensions 
$L_x,L_y$ and $L_z$. The Coulomb potential energy of this system with periodic
boundary condition in two directions and free boundary conditions in the third 
direction(let us say in the $z$ direction) can be written as

\begin{equation}
\label{eq01}
E=\frac{1}{2}\sum_{\bf n}' \sum_{i,j=1}^{N} \frac{q_i q_j}{|{\bf r}_{ij} + {\bf n}|}
\end{equation}

where ${\bf r}_{ij}={\bf r}_i - {\bf r}_j$ and ${\bf n}=(n_x L_x,n_y L_y,0)$, 
with $n_x,n_y$ being integers. The prime on the outer sum denotes that for
${\bf n}=0$ the term $i=j$ has to be omitted.\\
In the Ewald-type methods the above very slowly converging sum over the
Coulomb potential function is split into two sums which converge exponentially
fast, one in real space and the other in the Fourier space. 
This splitting can be done by adding and subtracting a term
corresponding to the electrostatic energy of a system of smooth spherical
charge densities,$\rho_i({\bf r})$,  centered on the particle positions:
\begin{eqnarray}
E&=& \frac{1}{2}\sum_{\bf n}'\sum_{i,j=1}^{N}
\left[\frac{q_i q_j}{|{\bf r}_{ij} + {\bf n}|}
- \int \!\!\! \int \frac{\rho_i({\bf r}) \rho_j({\bf r'}+{\bf n})}{|{\bf r}-{\bf r'}|}
d{\bf r}  d{\bf r'}\right]  \nonumber \\
&& +\frac{1}{2} \sum_{\bf n}
\sum_{i,j=1}^{N}\int \!\!\! \int
\frac{\rho_i({\bf r}) \rho_j({\bf r'}+ {\bf n})}{|{\bf r}-{\bf r'}|} d{\bf r}  d{\bf r'}  \nonumber \\
&& -\frac{1}{2} \sum_{i=1}^{N}\int \!\!\! \int
\frac{\rho_i({\bf r}) \rho_i({\bf r'})}{|{\bf r}-{\bf r'}|} d{\bf r}  d{\bf r'}
\label{eq02}
\end{eqnarray}
The aim of the last term is to subtract the self energy for ${\bf n}=0$ and $i=j$ which is
included in the second term. \\
Even though Ewald-type methods allow for any choice of $\rho_i(r)$, it was 
noted in Refs.[\onlinecite{cite04,cite05}] that Gaussians are virtually optimal in practice.
Choosing
$\rho_i(r)$ to be a Gaussian function 
\begin{equation}
\label{eq24}
\rho_i({\bf r})=\frac{q_i}{(\alpha^2 \pi)^{\frac{3}{2}}}
\; \; \exp\left[-\frac{|{\bf r}-{\bf r}_i|^2}{\alpha^2}\right]
\end{equation}
leads to a well-known formula for the first and the third term in Eq.(\ref{eq02}).
\begin{eqnarray}
\label{eq04}
E=&& \frac{1}{2}\sum_{\bf n}' \sum_{i,j=1}^{N}
\frac{q_i q_j \; \; \rm{erfc}\, \left[\frac{|{\bf r}_{ij} + {\bf n}|}{\alpha\sqrt{2}}\right]}
{|{\bf r}_{ij} + {\bf n}|}  + \nonumber \\
&& +\frac{1}{2} \sum_{\bf n}
\sum_{i,j=1}^{N}\int \!\!\! \int
\frac{\rho_i({\bf r}) \rho_j({\bf r'}+ {\bf n})}{|{\bf r}-{\bf r'}|} d{\bf r}  d{\bf r'}  \nonumber \\
&& -\frac{1}{\alpha \sqrt{2 \pi}} \sum_{i=1}^{N} q_i^2
\end{eqnarray}

Obviously, the calculation of the third term is trivial. Since the interaction
in the first term is decaying exponentially it can be made of finite range by 
introducing a cut-off.
The error resulting from the cut-off is then also exponentially small and 
the short range term can be calculated with linear scaling. We have calculated 
the short range part and also the contribution of forces from long range as it is 
described in Ref.[\onlinecite{cite04}]

The major difficulty is the calculation of the second term. 
A method to solve the Poisson's equation under 2DP1DF boundary conditions has 
recently been put forward by L. Genovese~\cite{cite18}.
Our approach is similar. As in Ref~[\onlinecite{cite18}] we use plane waves\cite{cite27} to represent the charge density in the 
periodic directions. 
Whereas Genovese et al used scaling functions as the basis in the non-periodic direction, 
we use finite elements for that purpose. 
Scaling functions are presumably the optimal choice in the context of electronic structure calculations 
where the charge density is given on a numerical grid. In our case the charge distribution 
is a sum over smooth Gaussians that can easily be represented by our mixed basis set of plane waves and 
finite elements. As will be seen we can avoid storing any kernel if we solve 
a differential equation along the z-axis instead of solving an integral equation. 
We will use a family of finite elements that allows to solve  
the linear system of equations resulting from the differential equation very efficiently. 

\subsection{Calculating the long range part}
The second term in Eq.(\ref{eq04}), can be written as

\begin{eqnarray}
\label{eq19}
E_{long}=\frac{1}{2} \int_{\Re^{3}} \rho^{(N)}({\bf r}) V({\bf r}) d{\bf r}
\end{eqnarray}
where
\begin{subequations}
\begin{eqnarray}
\rho^{(N)}({\bf r})&:=&\sum_{i=1}^{N} \rho_i({\bf r}) \\
V({\bf r})&:=&\int_{\Re^{3}} \frac{\rho({\bf r}')}{|{\bf r}-{\bf r}'|} d{\bf r}' \label{eq03} \\
\rho({\bf r})&:=&\sum_{\bf n} \sum_{j=1}^{N} \rho_j({\bf r}+ {\bf n}) 
\end{eqnarray}
\end{subequations}
We consider a system with a charge density 
that is only localized in the non-periodic direction, in our notation $z$; 
$\rho(x,y,z)=0 \ \forall (x,y,z) \in \Re^{3} \mid z \notin [z_l,z_u]$.
We define the cell containing the continuous charge density as:
\[\mathcal{V}:=[0,L_x]\otimes [0,L_y]\otimes [z_l,z_u] \]
In our case the length of $\mathcal{V}$ in $z$ direction $z_u-z_l$ 
is $L_z$ plus twice the cut-off for Gaussians. Since 
$\rho({\bf r})$ is periodic in $x$ and $y$ direction, 
$V({\bf r})$ is periodic too, so we can rewrite Eq.(\ref{eq19}) as:
\begin{eqnarray}
\label{eq20}
E_{long}=\frac{1}{2} \int_{\mathcal{V}} \rho({\bf r}) V({\bf r}) d{\bf r}
\end{eqnarray}

and $V({\bf r})$ can be calculated in an alternative way to Eq. (\ref{eq03}).
It can be considered as the solution of Poisson's equation with 2DP1DF BC:
\begin{eqnarray}
\label{eq07}
\nabla^2 V({\bf r})=-4\pi\rho({\bf r})
\end{eqnarray}

The charge density and the potential are periodic in 
$x$ and $y$ directions. Hence we can write the potential and
the charge density in terms of Fourier series:
\begin{subequations}
\begin{eqnarray}
V(x,y,z)&\hspace{-0.15cm}=\hspace{-0.25cm}&\sum_{k,l=-\infty}^{\infty} c_{kl}(z) \exp\left[2i\pi(\frac{k \, x}{L_x}+\frac{l \, y}{L_y})\right]
\label{eq05} \\
\rho(x,y,z)&\hspace{-0.15cm}=\hspace{-0.25cm}&\sum_{k,l=-\infty}^{\infty} \frac{\eta_{kl}(z)}{-4\pi} \exp\left[2i\pi(\frac{k\,x}{L_x}+\frac{l\,y}{L_y})\right]
\label{eq06}
\end{eqnarray}
\end{subequations}

Inserting Eqs.(\ref{eq05}) and (\ref{eq06}) in Eq.(\ref{eq07}) yields:
\begin{equation}
\label{eq13}
\fbox{$\left( \frac{d^2}{dz^2} - \gamma_{kl}^2\right) c_{kl}(z) =\eta_{kl}(z)$} 
\end{equation}
\begin{equation}
\gamma_{kl}:=2\pi\sqrt{\frac{k^2}{L_x^2}+\frac{l^2}{L_y^2}} \nonumber 
\end{equation}
\begin{eqnarray}
\eta_{kl}(z)=&& \frac{-4\pi}{L_x L_y} \int_{0}^{L_x} \int_{0}^{L_y} \rho(x,y,z) \nonumber\\
&& \times\exp\left[-2i\pi(\frac{k\,x}{L_x}+\frac{l\,y}{L_y})\right] dx dy
\end{eqnarray}
To solve the differential equation (\ref{eq13}) one needs to have 
boundary conditions at $z\rightarrow \pm\infty$ for $c_{kl}(z)$. 
The potential obtained by solving Poisson's equation should be the 
same as the one in Eq.~(\ref{eq03}). Hence we derive the boundary condition in the non-periodic 
direction from Eq.~(\ref{eq03}). By performing the Taylor expansion of 
$\frac{1}{|\vec{r}-\vec{r}'|}$ about $z'=0$ in 
the integral expression of Eq.~(\ref{eq03}) for the exact potential $V(x,y,z)$ arising from 
our periodic charge distribution $\rho({\bf r})$ 

\begin{eqnarray}
V(x,y,z) =&&\int_{z_l}^{z_u}\int_{-\infty}^{\infty} \int_{-\infty}^{\infty} dx' dy' dz' \frac{1}{|\vec{r}-\vec{r}'|} \nonumber \\
&& \hspace{-1.6cm}\times \sum_{k,l=-\infty}^{\infty} \frac{\eta_{kl}(z')}{-4\pi}\exp\left[2\pi i(\frac{kx'}{L_x}+\frac{ly'}{L_y}) \right] 
\end{eqnarray}
one can show that 
$ V(x,y,z\rightarrow \pm\infty) = \mp\beta$ 
where $\beta$ is proportional to the dipole moment of the charge distribution along the z direction
\begin{eqnarray}
\beta= \frac{1}{2} \int_{z_l}^{z_u}\eta_{00}(z') z' dz' \label{eq22}
\end{eqnarray}
For the Gaussian charge distributions given by Eq.~(\ref{eq24}) the above integral can 
be calculated analytically and $\beta$ is calculated exactly.
\begin{equation}
\beta= \frac{-2\pi}{L_x L_y} \sum_{i=1}^{N} q_i z_i
\end{equation}

This boundary condition for the potential gives the following conditions for the $\gamma$'s.
\begin{itemize}
\item $\gamma=\gamma_{00}=0 \Rightarrow \frac{d^2}{dz^2} c_{00}(z) =\eta_{00}(z)$  \quad  We solve
      this differential equation with boundary condition 
      $c_{00}(z\rightarrow \pm\infty)=\mp \beta$
\item $\gamma=\gamma_{kl}\neq0 \Rightarrow \left( \frac{d^2}{dz^2} - \gamma_{kl}^2\right) c_{kl}(z) =\eta_{kl}(z)$
      \quad For all of these differential equations we have to impose BC of the form
      $c_{kl}(z\rightarrow\pm\infty)=0$.
\end{itemize}
The solution for $c_{00}(z)$ is a linear function outside the interval $[z_l,z_u]$.
Since the boundary conditions are applied at infinity the linear term has to vanish 
and one has to satisfy Dirichlet BC 
for $c_{00}$, namely $c_{00}(z_u)=-\beta$ and $c_{00}(z_l)=\beta$.
For $|k|+|l|>0$, $c_{kl}(z)$ will have Robin BC as explained below. 
The potential is thus not modified if 
one takes for instance a computational box that is thicker in the $z$ direction than necessary. 
The thinnest possible box is the one that just includes the region where the charge is nonzero.
 
For $z \in  (-\infty,z_l]$ we have $\eta_{kl}(z)=0$ thus it yields 
\begin{equation}
c(z)=c(z_l) e^{\gamma_{kl}(z-z_l)} 
\end{equation}
Both $c(z)$ and its derivative must be continuous. So performing left 
differentiation at $z_l$ we get:
\begin{equation}
c'(z_l) - \gamma_{kl} c(z_l)=0
\end{equation}
With a similar procedure we obtain the BC at $z_u$:
\begin{equation}
c'(z_u) + \gamma_{kl} c(z_u)=0
\end{equation}
These BCs are in agreement with the BCs resulting from the Green functions 
in Ref.~[\onlinecite{cite18}]

\subsection{Solving the ordinary differential equation using 
the finite element method}

We recapitulate the procedure of solving the differential equation for the case $|k|+|l|>0$, i.e. $\gamma_{kl} \neq 0$, using the 
finite element method. For the case $k=l=0$ the approach is similar, with the only difference that the 
Dirichlet BC are used. The case $k=l=0$ can be found in many manuscripts and textbooks  on the finite element method 
e.g. Ref.~[\onlinecite{cite25}]. In particular our notation follows Ref.~[\onlinecite{cite25}].
Discretizing the differential equation with mentioned Robin BCs using the finite element 
method leads to a system of linear equations. The resulting matrix is a banded matrix for which the 
system of equations can be solved efficiently  
if high-order hierarchical piecewise polynomials are used as a basis  and if 
the degrees of freedom are decimated. 
This hierarchical finite element basis set leads to algebraic systems that are less susceptible to round-off 
error accumulation at high order than those produced by a Lagrange basis\cite{cite24}. 
We use linear hat functions as the linear hierarchical basis. For
higher order bases we exploit the method of Szab\'{o} and Babu\v{s}ka\cite{cite08} which relies on 
Legendre polynomials. Below we show the expansion of $c(z)$ in terms of the hat functions and the other 
higher order hierarchical piecewise polynomials on the interval $[z_{i-1},z_i]$:
\begin{equation} \label{eq25}
c(z)\approx c_{i-1} N_{-1}(\xi_i)+c_{i} N_{1}(\xi_i) +\sum_{j=2}^{p}c_{i,j} N_j(\xi_i) \: ,
\end{equation}
where $\xi_i=2(z-z_i)/h + 1; \, h=z_{i}-z_{i-1}$ and the functions $N_i(\xi)$ in the interval $[-1,1]$ are given by 
\begin{subequations} \label{eq23}
\begin{eqnarray}
N_{-1}(\xi)&=&\frac{1-\xi}{2} \hspace{1cm} N_{1}(\xi)=\frac{1+\xi}{2} \\
N_i(\xi)&=&\sqrt{\frac{2i-1}{2}} \int^{\xi}_{-1} P_{i-1}(\xi')d\xi', \ \ \ i\geq 2
\end{eqnarray}
\end{subequations}
These hierarchical bases have useful orthogonality properties that
lead to sparse and well-conditioned stiffness matrices.
Defining an operator $\mathcal{L}$ 
\begin{equation}
\mathcal{L}[c]:=c''(z)-\gamma^2 c(z)
\label{eq17}
\end{equation}
we can write our differential equation (\ref{eq13}) as
$$ \mathcal{L}[c]=\eta(z) $$
with boundary conditions
\begin{eqnarray}
\left\{
\begin{array}{c}
c'(z_l)-g c(z_l)=0 \\
c'(z_u)+g c(z_u)=0
\end{array}
\right.
\label{eq08}
\end{eqnarray}
The method of weighted residuals is used to construct a variational integral 
formulation of Eq.(\ref{eq17}) by multiplying 
with a test function $d(z)$ and integrating over $[z_l,z_u]$:
\begin{eqnarray}
\label{eq21}
(d,\mathcal{L}[c]-\eta)=0  \ \ \  \forall d\in H^1(z_l,z_u)
\end{eqnarray}
where $H^1$ is the Sobolev space.
We have introduced the $L^2$  inner product
\begin{eqnarray}
(d,c):=\int^{z_u}_{z_l} d(z) c(z) dz
\end{eqnarray}
Performing the integration by parts in Eq.(\ref{eq21}) and 
applying Robin BCs given in Eq.(\ref{eq08}) gives 
\begin{equation} \label{eq11}
A(d,c)=(d,\eta)+g d(z_l) c(z_l) + g d(z_u) c(z_u) 
\end{equation}
where
\begin{eqnarray}
A(d,c):=\int^{z_u}_{z_l}\left[ -d'(z) c'(z) -\gamma^2 d(z) c(z)\right] dz
\end{eqnarray}
{
\sloppy
Using the Galerkin approach and exploiting the decimation scheme, we
can construct a system of linear equations $B \vec{c}=\vec{b}$
where the elements of the vector $\vec{c}$ are the values of $c(z)$ at grid points. 
The detailed structure of this linear system of equations is given in the Appendix A.
}

\section{Numerical Result}

In this section we present the numerical results obtained 
for the Poisson's solver for continuous charge densities with 2DP1DF BC  
in stand alone mode and for our Ewald-like method for point 
particles interacting by Coulombic potential with 2DP1DF BC. 
We also show numerical evidence for the conservation 
of energy in molecular dynamics simulation of a system composed of sodium 
chloride atoms. \par

\subsection{\label{sec:sec1}Numerical results for the Poisson solver}

\begin{figure}
\includegraphics[height=50mm]{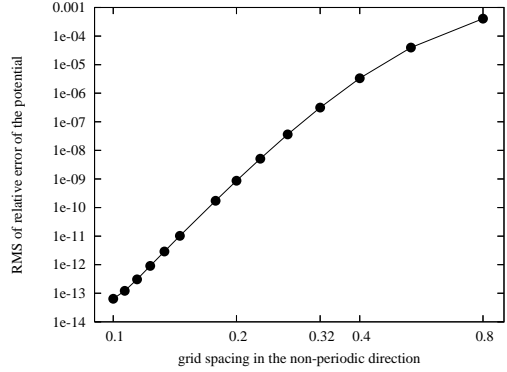}
\caption{\label{fig:fig1} RMS of relative error for the potential 
given in Sec.~\ref{sec:sec1} with $a=10, b=10, c=1$. On this double logarithmic plot the 
curve has an asymptotic slope of 14 and machine precision can be reached.}
\end{figure}

Our method has an algebraic convergence rate in the non-periodic 
direction and a faster exponential convergence rate in the periodic 
directions, respectively due to the finite element polynomial 
bases and to the plane wave representation. In Fig.~\ref{fig:fig1} 
we show the convergence rate in non-periodic direction with 7-th order finite elements 
(p=7 in Eq.~(\ref{eq25})). 
For our test, the starting point was the potential rather than the charge 
density, since the charge density can be obtained analytically from the potential 
by simple differentiation. 
Our test potential had the form 
$\phi({\bf r})=\sin(a \sin(\frac{2 \pi x}{L_x})) \sin(b \sin(\frac{2 \pi y}{L_y})) \exp(-\frac{z^2}{c^2})$.

\subsection{Numerical results for point particles}

\begin{figure}
\includegraphics[height=50mm]{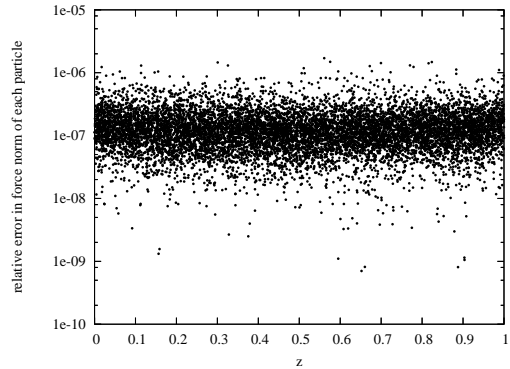}
\caption{\label{fig:fig2} Relative error distribution of force norm 
on each particle along z-axis for 100 random systems with 100 particles.}
\end{figure}

In this section we give the numerical results of our 
implementation of the presented method for point particles. 
Since MMM2D is known to be highly accurate we use it as reference in 
this section. First we want to demonstrate that error distribution 
along the non-periodic direction is uniform unlike in the 3D periodic 
methods with correction terms\cite{cite07,cite15,cite16}. To 
this aim 100 particles were put randomly in a unit cubic cell 
and the program was run 100 times each time with different random 
positions. Results of the relative error of forces exerted on 
each particle are plotted in Fig.~\ref{fig:fig2}. 

\begin{figure}
\centering
\includegraphics[width=80mm]{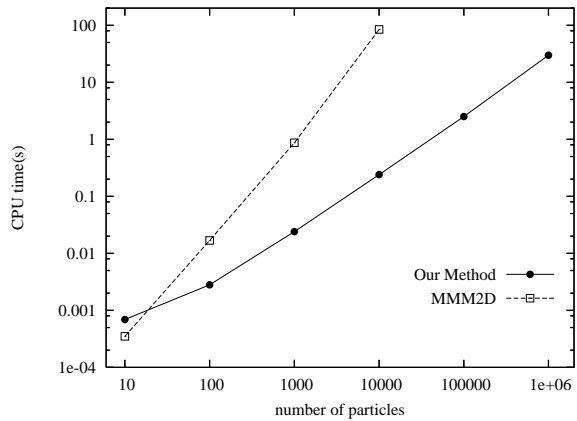}
\caption{\label{fig:fig3} CPU time of one time evaluation of 
forces on particles and potential energy with our method(solid curve) 
and MMM2D method(dashed curve).}
\end{figure}

In Fig.~\ref{fig:fig3} we show the theoretical scaling $\mathcal{O}(N \log(N))$ 
can be achieved in practice. The crossover with MMM2D for a moderate accuracy of 
$10^{-4}$ in RMS relative error of forces is found to be less than 
20 particles. Both programs were run in AMD Opteron 2400 MHz.
The degree of the finite elements is a parameter that can be optimized to 
obtain the smallest possible CPU time for a fixed accuracy. 
For high accuracies higher degrees are recommended.
The CPU time for the calculation of the forces dominates in our method over the time 
needed to calculate the energy.


\subsection{Energy Conservation}

\begin{figure}
\centering
\includegraphics[width=80mm]{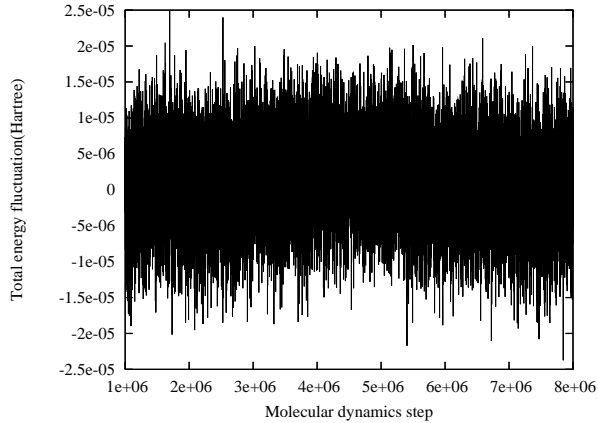}
\caption{\label{fig:fig4} The total energy fluctuations calculated with 
our method.}
\end{figure}

Energy conservation is of great importance in molecular dynamics simulations. 
In order to investigate energy conservation in a real simulation, we performed a very long
(8 nano second) molcular dynamics simulation of a sodium chloride system 
containing 1000 particles. The velocity Verlet algorithm with a time step 
of 50 atomic units is used to update the particle positions and velocities. 
The short range interactions are obtained from the 
Born-Mayer-Huggins-Fumi-Tosi\cite{cite22} (BMHFT) rigid-ion potential, 
with the parameters of Ref.[\onlinecite{cite23}]. The shortest 
oscillation period was of the order of 3000 atomic units e.i. 60 
molecular dynamics steps. After an equilibration for $1\times 10^6$ steps, 
$7\times 10^6$ steps were performed during which the total energy and potential 
energy were monitored. The fluctuation of the total 
energy, shown in Fig.~\ref{fig:fig4}, has an oscillation amplitude of about $2.5\times 10^{-5}$, 
while the amplitude of the potential energy oscillation was 3 orders of magnitude 
larger. The total energy was thus conserved very well. \par

\section{Conclusion}
In this manuscript we presented a method to solve 
Poisson's equation for smooth charge densities with periodic 
boundary condition in two directions and finite in the third one. 
It is very efficient for smooth charge densities and it does not require 
much memory.  The resulting error distribution is uniform over the entire 
simulation cell. Our method is based on plane wave representation in the periodic 
directions and finite elements in the non-periodic direction. 
Based on this method we can then calculate electrostatic energy and 
forces of particles interacting by Coulombic potential with high accuracy and an $N \log(N)$ scaling.
The method satisfies intrinsically and without any approximations the boundary 
conditions approriate for surface problems.
It is best suited for a moderate 
number of particles in between  $10^2-10^6$. 
The method is expected to be suitable for an efficient parallelization 
since the time dominating parts are only loosely coupled.

\begin{acknowledgments}
One of the authors would like to thank M. J. Rayson for 
valuable and helpful discussions. This work has been supported by the 
Swiss National Science Foundation and the Swiss National Center of Competence in Research(NCCR) on Nanoscale Science. 
\end{acknowledgments}

\appendix

\section{Appendix}

We consider a uniform grid on the interval $[z_l,z_u]$ with $N+1$ nodes $\{z_0,z_1,\dots,z_N\}$  while  
$z_0=z_l$ and $z_N=z_u$. The interval is thus divided 
into $N$ equally spaced subintervals(elements). The functions $d(z)$ and $c(z)$ 
are replaced by the approximate functions $D(z)$ and $C(z)$ which 
are expanded in the basis of Eqs.~(\ref{eq23}) on each subinterval. We use the Galerkin approach 
in which the same bases are used for the expansion of both $D(z)$ and $C(z)$. 
Our bases are a combination of the hat function $\phi^v(z)$ centered at nodes 
\begin{eqnarray}
\phi^v_j(z)&=&\left\{ 
\begin{array}{lr}
(z_{j+1}-z)/h, & z\in [z_j,z_{j+1}) \\
(z-z_{j-1})/h, & z\in [z_{j-1},z_{j}) \\
0              & \rm{otherwise}
\end{array} \right. \\
\end{eqnarray}
and hierarchical polynomials\cite{cite08} $\phi^m(z)$ 
\begin{eqnarray}
\phi^m_{j,i}(z)&=&\left\{
\begin{array}{lr}
N_i(2\frac{z-z_{j}}{h}+1), & z\in [z_{j-1},z_{j}] \\
0              & \rm{otherwise}
\end{array} \right.
\end{eqnarray}
localized within the individual elements. $N_i$ are given in canonical coordinates in Eqs.(\ref{eq23}).
Finally $C(z)$ and $D(z)$ within the element $[z_{j-1},z_j]$ 
will be:
\begin{subequations}
\begin{eqnarray}\label{eq09}
&&\hspace{-0.7cm} C(z)=c_{j-1} \phi^v_{j-1}(z) + c_j \phi^v_j(z) + \sum^{p}_{i=2} c_{j.i} \phi^m_{j,i}(z) \\
\label{eq10}
&&\hspace{-0.7cm}D(z)=d_{j-1} \phi^v_{j-1}(z) + d_j \phi^v_j(z) + \sum^{p}_{i=2} d_{j,i} \phi^m_{j,i}(z)
\end{eqnarray}
\end{subequations}

 Note that 
because $\phi^m_{j,i}(z)$ vanishes at all nodes we obtain $c_j=C(z_j)$. Replacing 
the approximate functions from Eq.(\ref{eq09}) and Eq.(\ref{eq10}) 
in equation (\ref{eq11}) gives 
\begin{eqnarray}
\label{eq18}
\sum^{N}_{j=1}[A_j(D,C)-(D,\eta)_j]=g d_0 c_0 + g d_N c_N
\end{eqnarray}
We split  $A_j(D,C)$ as 
\begin{eqnarray}
A_j(D,C)=A^S_j(D,C) + A^M_j(D,C)
\end{eqnarray}
where
\begin{eqnarray}
A^S_j(D,C)&:=&-\int^{z_j}_{z_{j-1}} D'(z) C'(z) dz     \\
A^M_j(D,C)&:=&-\int^{z_j}_{z_{j-1}} \gamma^2 D(z) C(z) dz   \\
(D,\eta)_j&:=&\int^{z_j}_{z_{j-1}} D(z) \eta(z) dz
\end{eqnarray}

$C(z)$ within an element is:
\begin{equation}
C(z)=\vec{\phi}_j^T(z) \vec{c}_j  ~~~~~  z\in [z_{j-1},z_j]
\end{equation}
where $\vec{c}_j$ and $\vec{\phi}_j(z)$ are vectors with $p+1$ elements:
\begin{eqnarray}
\vec{c}_j&:=&[c_{j-1},c_j,c_{j,2},\dots,c_{j,p}]^T  \\
\vec{\phi}_j(z)&:=&[\phi^v_{j-1}(z),\phi^v_{j}(z),\phi^m_{j,2}(z),\dots,\phi^m_{j,p}(z)]^T
\end{eqnarray}
Then
\begin{eqnarray}
A^S_j(D,C)&=&\vec{d}^T_j K_j \vec{c}_j \\
A^M_j(D,C)&=&\vec{d}^T_j M_j \vec{c}_j 
\end{eqnarray}
where
\begin{eqnarray}
K_j&:=&-\int^{z_i}_{z_{i-1}} \frac{d \vec{\phi}_j}{dz} ~ \frac{d \vec{\phi}^T_j}{dz} \\
M_j&:=&-\int^{z_i}_{z_{i-1}} \gamma^2 \vec{\phi}_j \ \vec{\phi}^T_j
\end{eqnarray}
The $(p+1) \times (p+1)$ matrix $K_j$ is called the element stiffness 
matrix and the $(p+1) \times (p+1)$ matrix $M_j$ is called the element 
mass matrix. Although the element index $j$ is present in the definition 
of $K_j$ and $M_j$, in our case of uniform grid spacing these matrices do not 
depend on $j$. By performing the summation 
$\sum_{j=1}^{N} A^M_j$ and $\sum_{j=1}^{N} A^S_j$, we build up the global 
mass matrix and the global stiffness matrix. We arrange the order of elements of these matrices as:
\begin{eqnarray} \label{eq27}
\vec{c}&:=&
\left[
\begin{array}{l}
\vec{c}_L \\
\vec{c}_Q
\end{array}
\right] \\
\vec{c}_L&:=&[c_0,c_1,\dots,c_N]^T  \\
\vec{c}_Q&:=&[c_{1,2},\dots,c_{1,p},\dots,c_{N,2},\dots,c_{N,p}]^T  
\end{eqnarray}
\begin{eqnarray}
K&=&
\left[
\begin{array}{cc}
K_L  &  0    \\
0    & K_Q
\end{array}
\right] \\
M&=&
\left[
\begin{array}{cc}
M_L  &  M_{LQ}    \\
M^T_{LQ}  &  M_Q
\end{array}
\right]
\end{eqnarray}
The second term of the summand in Eq.(\ref{eq18}) should be 
calculated approximately because only the values of $\eta(z)$ 
on the nodes are available:
\begin{equation}
(D,\eta)_j=\vec{d}^T_j \vec{I}_j  
\end{equation}
where
\begin{equation}
\vec{I}_j:=\int^{z_j}_{z_{j-1}} \vec{\phi}_j(z) \eta(z) dz 
\end{equation}
Interpolating integration is appropriate to calculate the 
above integral by fitting a polynomial of degree $d\geq 2p$ 
to the nodes of element $[z_{j-1},z_j]$ and its neighboring nodes:
\begin{equation}
(\vec{I}_j)_i=\sum^{p-1}_{k=-p} w^i_k \eta_{j+k}
\end{equation}
Recall that our charge density is localized within the interval $[z_l,z_u]$ and 
it smoothly tends to zero at the edges. Therefore it is appropriate to zero pad 
the ends of the $\eta(z)$. 
The coefficients $w^i_k$ are weights from high-order interpolation. Building up 
the global matrices yields:
\begin{equation}
(D,\eta)=\vec{d}^T \vec{I} 
\end{equation}
where the order of elements of $\vec{I}$ is the same as in Eq.~(\ref{eq27}),
\begin{eqnarray}
\vec{I}&:=&
\left[
\begin{array}{l}
\vec{I}_L \\
\vec{I}_Q
\end{array}
\right] \\
\vec{I}_L&:=&[I_0,I_1,\dots,I_N]^T  \\
\vec{I}_Q&:=&[I_{1,2},\dots,I_{1,p},\dots,I_{N,2},\dots,I_{N,p}]^T  
\end{eqnarray}
Finally by adding the right-hand-side of Eq.(\ref{eq18}) to the global matrices yields:
\begin{equation}
\label{eq15}
\left[
\begin{array}{cc}
P_L  &  M_{LQ}    \\
M^T_{LQ}  &  P_Q
\end{array}
\right]
\left[
\begin{array}{l}
\vec{c}_L \\
\vec{c}_Q
\end{array}
\right]
=
\left[
\begin{array}{l}
\vec{I}_L \\
\vec{I}_Q
\end{array}
\right]    \\
\end{equation}
where $M_{LQ}$ is a sparse $(N+1)\times N(p-1)$ matrix, 
\begin{eqnarray}
P_Q&:=&K_Q + M_Q 
\end{eqnarray}
is a $N(p-1)\times N(p-1)$ block-diagonal matrix, 
\begin{eqnarray}
P_L&:=&K_L + M_L + g e_0 e_0^T +  g e_N e_N^T 
\end{eqnarray}
is a tridiagonal $(N+1)\times (N+1)$ matrix, and 
\begin{eqnarray}
e_0&:=&[1,0,\dots,0]^T \\
e_N&:=&[0,\dots,0,1]^T 
\end{eqnarray}

Multiplying the matrix in Eq.(\ref{eq15}) 
and eliminating $\vec{c}_Q$ in the system of linear equations 
yields:
\begin{equation}
\left[
P_L -M_{LQ} P^{-1}_Q M^T_{LQ}
\right]
\vec{c}_L=\vec{I}_L - M_{LQ} P^{-1}_Q \vec{I}_Q
\end{equation}

Finally we obtain our system of linear equations:
\begin{equation}
B\vec{c}_L=\vec{b}
\end{equation}

where the matrix $B$ and the vector $\vec{b}$ are 
\begin{eqnarray}
B&:=&P_L -M_{LQ} P^{-1}_Q M^T_{LQ} \\
\vec{b}&:=&\vec{I}_L - M_{LQ} P^{-1}_Q \vec{I}_Q 
\end{eqnarray}
It turns out that in the general case the matrix $B$ is symmetric tridiagonal of dimension $(N+1)\times (N+1)$. The proof 
for the tridiagonality of matrix $B$ can be found in the context of block cyclic reduction\cite{cite26}. Note that elements of the 
vector $\vec{c}_L$ are the values of $C(z)$ at the grid points. Therefore by 
solving a system of linear equations, which has a tridiagonal matrix, we can 
find the values of $C(z)$ at the grid points. Instead of using finite element method, we could 
have used finite differences to solve Eq.~(\ref{eq13}). Although calculating the right-hand-side $\vec{b}$ 
is computationally more expensive in our approach than in the finite difference method, the whole process of 
solving the system of linear equations is less expensive because the factorization 
of the tridiagonal matrix can be done fast. 

\bibliography{eepp2dp1df}

\end{document}